%Time evolution -- entropy.tex

\magnification=\magstep1
\overfullrule=0pt

\def\frac#1#2{{#1\over #2}}           % fractions in vertical
\let\bm=\bf   \let\boldnabla=\nabla

\font\chapfont=cmbx10 scaled \magstep2

\font\header=cmcsc10
       \font\ninerm=cmr9
       
   \font\nineit=cmti9
\font\dots=cmbx10

\def\nh{\noindent\hang}
\def\eq#1{\eqno\hbox{(#1)}}

% Repeated author entry.
\def\r{\vrule height0.6pt depth0pt width1.5cm \  }

\font\bigmath=cmsy10 scaled \magstep1
\def\bigsigma{\raise 1pt\hbox{\bigmath$\sigma$}}
\def\biggamma{\raise 1.5pt\hbox{\bigmath$\gamma$}}

%% ETJ redefinitions of \footnote
\def\fnd#1{{\parindent=10pt\baselineskip=12 true pt\mathsurround=0pt%
    \def\it{\nineit}\footnote{{$^\dagger$}}{\ninerm {#1}}}} %
%          ddagger
  %              star
  %          hand numbered

%Subsection Heading
\outer\def\heading{\bigbreak\bgroup \let \\=\cr \tabskip
      \centering \halign to \hsize\bgroup \header
                 \hfill\ignorespaces## \unskip\hfill\cr}
     \def\endheading{\cr\egroup\egroup \nobreak
                        \medskip\noindent}
%Example: \heading
%         This is how we ... \\... include extra stuff
%         \endheading
%  Note that \\ splits the heading into 2 centered lines.

\centerline{\chapfont Time Evolution In Macroscopic Systems.}\par
\centerline{\chapfont II: The Entropy} \bigskip
\centerline{\dots W.T. Grandy, Jr.} \bigskip
\centerline{\dots Department of Physics \& Astronomy, University of Wyoming}\smallskip
\centerline{\dots Laramie, Wyoming 82071}
\bigskip

\noindent {\bf Abstract.} The concept of entropy in nonequilibrium macroscopic systems  is 
investigated in the light of an extended equation of motion for the density matrix obtained 
in a previous study. It is found that a time-dependent information entropy can be defined 
unambiguously, but it is the time derivative or entropy production that governs ongoing 
processes in these systems. The differences in physical interpretation and thermodynamic role
of entropy in equilibrium and nonequilibrium systems is emphasized and the observable aspects 
of entropy production are noted. A basis for nonequilibrium thermodynamics is also outlined. 
\bigskip

\noindent {\bf 1. Introduction}\medskip

The empirical statement of the Second Law of thermodynamics by Clausius (1865) is
$$
S(initial)\le S(final)\,, \eq1
$$
where $S$ is the total entropy of everything taking part in the process under consideration,
and the entropy for a single {\it closed} system is defined to within an additive constant by
$$
S(2)-S(1) =\int_1^2 \frac{dQ}{T}=\int_1^2 C(T)\frac{dT}{T}\,, \eq2
$$
where $C(T)$ is a heat capacity.
The integral in (2) is to be taken over a reversible path, a locus of
thermal equilibrium states connecting the macroscopic states $1$ and $2$,
which is necessary because the absolute temperature $T$ is not defined for other than
equilibrium states; $dQ$ represents the net thermal energy added to or taken
from the system in the process.  As a consequence, entropy is defined in
classical thermodynamics {\it only} for states of thermal equilibrium.
Equation (1) states that in the change from one state of thermal equilibrium to another
along a reversible path the total entropy of all bodies involved cannot
decrease; if it increases, the process is irreversible. That is, the integral provides
a lower bound on the change in entropy. This phenomenological entropy is to be found from experimental measurements with
calorimeters and thermometers, so that by construction it is a function only of the macroscopic
parameters defining the macroscopic state of a system, $S(V,T,N)$, say, where $V$ and $N$ are the system
volume and particle number, respectively. It makes no reference to microscopic variables or probabilities,
nor can any explicit time dependence be justified in the context of classical
thermodynamics. Equation (1) is a statement of macroscopic phenomenology
that cannot be proved true solely as a consequence of the microscopic dynamical
laws of physics, as appreciated already by Boltzmann (1895): ``The Second Law can never be proved
mathematically by means of the equations of dynamics alone." (Nor, for that matter, can the First Law!)

Theoretical definitions of entropy were first given in the context of statistical mechanics by Boltzmann and 
Gibbs, and these efforts culminated in the formal definition of equilibrium ultimately given by Gibbs (1902) in terms
of his variational principle. In Part I (Grandy, 2003, preceding paper\fnd{Equation (n) of that paper will
be denoted here by (I--n).})
we observed that the latter is a special case of a more general principle of maximum information
entropy (PME), and in equilibrium it is that maximum subject to macroscopic constraints that is identified with
the experimental entropy of (2). One of the dominant concerns in statistical mechanics and thermodynamics has long
been that of extending these notions unambiguously to nonequilibrium phenomena and irreversible processes, and it is that
issue we shall address in this work.

How is one to define a time-dependent $S(t)$ for nonequilibrium states? Do there even exist sensible physical definitions
of experimental and theoretical `entropy' analogous to those describing equilibrium states? Other than
$S(t)$ and the density matrix $\rho(t)$, what other key parameters might be essential to a complete description of 
nonequilibrium? These and other
questions have been debated, sometimes heatedly, for over a century without any broad concensus having been reached;
perhaps a first step toward clarifying the issue should be to understand the source of the differences of opinion
that lead to so many different points of view.
One problem, of course, is a lack of experimental guidance in determining those features of the phenomena that
are really of fundamental importance, and associated with this has been the necessary restriction of theories to
linear departures from equilibrium, owing to enormous calculational difficulties with the appropriate nonlinear
forms. What happens, then, is that many theoretical descriptions of nonequilibrium systems tend to predict similar
results in the linear domain and there is little to distinguish any fundamental differences that get to deeper matters.

In view of these obstacles it may be useful to look at the problem from a different perspective. Such sharp
disagreements would seem to arise from different hidden premises in various approaches to nonequilibrium statistical
mechanics, and we suggest here that these have much to do with differing views of the underlying probability
theory and its precise role. We initiated an examination of this point in I, which culminated in the expression
(I--40) as the appropriate form of the equation of motion for the density matrix. Our purpose here is to apply
the implications of that result to a further study of time varying macroscopic systems.

As a preliminary step it might be helpful to note some overall features of entropy and nonequilibrium processes
that have to be considered in any approach to the problem of generalizing $S$. Suppose we prepare a system in a
nonequilibrium state by applying an external force of some kind that substantially perturbs the equilibrium system,
possibly increasing its energy and adding matter to it, for example.
This state is defined by removing the external source at time $t=t_0$, and at that instant it is described by a
density matrix $\rho(t_0)$. Whether or not we can define a physical entropy at that time, we can certainly compute
the information entropy of that nonequilibrium state as $S_I(t_0)=-k{\rm Tr}[\rho(t_0)\ln\rho(t_0)]$. Because the
system is now isolated, $\rho(t)$ can only evolve from $\rho(t_0)$ by unitary transformation and $S_I$ remains constant into the
future.\fnd{Because there are numerous `entropies' defined in different contexts, we shall denote the experimental
equilibrium entropy of Clausius as $S$ without further embellishments, such as subscripts.} What happens next?

At the cutoff $t=t_0$ the entropy $S_I(t_0)$ refers only to the nonequilibrium state at that time. In the absence
of any other external influences we expect the system to relax into a new equilibrium state, for no other reason than
it is the one that can be realized in the overwhelmingly greatest number of ways subject to the appropriate macroscopic
constraints. The interesting thing is that those constraints are already fixed once the external sources are removed,
so that the total energy, particle number, volume, etc. at $t_0$ are now determined for $t>t_0$. The entropy of the
final equilibrium state is definitely {\it not} $S_I(t_0)$, but it is in principle known well before equilibrium is reached:
it is the maximum of the information entropy subject to constraints provided by the values of those thermodynamic
variables at $t=t_0$. We may or may not know these values, of course, although a proper theory might predict them;
but once the final macrostate is
established they can be measured and a new $\rho_f$ calculated by means of the PME, and hence a new entropy predicted
for comparison with the experimental form of Clausius; indeed, in equilibrium the Clausius entropy (2) is an upper bound for $S_I$.
Thus, in this relaxation mode we do not see a nice,
continuous, monotonically increasing entropy function that can be followed into the equilibrium state; but that's not
too surprising, given that we know $\rho(t_0)$ cannot evolve unitarly into $\rho_f$. There remains a significant 
dynamical evolution during this relaxation period, but it is primarily on a microscopic level; its macroscopic 
manifestation is to be found in the relaxation time, and the possible observation of decaying currents.
One  thing we might
compute and measure in this mode is that relaxation time, which does not have an immediate or necessary connection
with entropy. (There may, however, exist a `relaxation entropy' associated with the redistribution of energy during
the relaxation period, say.)

Ironically, the equilibrium state described so well by classical thermodynamics is essentially a dead end; it is a {\it singular
limit} in the sense discussed by Berry (2002). Equilibrium is actually a very special, ideal state, for most
systems are not usually in equilibrium, at least not completely. As external influences, and therefore time
variations, become smaller, the system still remains in a nonequilibrium state evolving in time. In the limit
there is a discontinuous {\it qualitative} change in the macroscopic system and its description. That is, there is
no longer either an `arrow of time' or a past history\fnd{Requiring the equilibrium system to have no `memory' of its past
precludes `mysterious' effects such as those caused by spin echos.}, and the main role of the theory is to compare
neighboring states of thermal equilibrium without regard for how those states might have been prepared.

It has long been understood that entropy is not a property of a physical system {\it per se}, but of the
thermodynamic system describing it, and the latter is defined by the macroscopic constraints imposed. The above
remarks, however, lead us to view entropy more as a property of the macrostate, or of the processes taking place
in a system. In the equilibrium state these distinctions are blurred, because the thermodynamic system and
the macrostate appear to be one and the same thing and there are no time-dependent macroscopic processes. 
It will be our goals in the following paragraphs to clarify these comments,
as well as to provide both an unambiguous definition of entropy in nonequilibrium systems, and to
understand the possibly very different roles that the entropy concept plays in the two states. \bigskip

\noindent {\bf 2. Some Preliminary Extensions of the Equilibrium Theory}\medskip

In I we briefly outlined the variational method of constructing an initial density matrix that could then evolve 
in time via the appropriate equations of motion. A principal application of that construction is to
equilibrium systems, in which case the quantum form of (I--14) becomes
$$
\rho_0=\frac1{Z}e^{-\beta H}\,, \qquad Z(\beta)={\rm Tr}e^{-\beta H}\,, \eq3
$$
where $H$ is the system Hamiltonian and $\beta=(kT)^{-1}$.
But, if there is no restriction to constants of the motion,
the resulting state described by (3) could just as well be one of nonequilibrium based
on information at some particular time. Data given only at a single point in space and time, however, can hardly
serve to characterize a system whose properties are varying over a space-time region, so a first generalization
of the technique is to information available over such regions.
Thus, the main task in this scenario is to gather information that varies in both space and time
and incorporate it into a density matrix describing a nonequilibrium state. Given an arbitrary but definite
thermokinetic history, we can look for what general behavior of the system can be deduced from just this.
The essential aspects of this approach were first expounded by Jaynes (1963, 1967, 1979).

To illustrate the method of information gathering, consider a system with a fixed time-independent
Hamiltonian and suppose the data to be given over a
space-time region $R({\bm x},t)$ in the form of an expectation value of a Heisenberg operator
$F({\bm x},t)$.
We are reminded that the full equation of motion for such operators, if they
are also explicitly time varying, is
$$
i\hbar{\dot F}=[F,H]+\partial_t F\,. \eq4
$$
When the data vary continuously over $R$
their sum becomes an integral and there is a distinct Lagrange multiplier for each space-time point.
Maximization of the entropy subject to the constraint provided by that information leads to a
density matrix describing this macrostate:
$$
\rho =\frac1{Z}\exp\left[-\int_R\lambda({\bm x},t)F({\bm x},t)\,d^3x\,dt\right]\,, \eq5
$$
where
$$
Z[\lambda({\bm x},t)]={\rm Tr}\exp\left[-\int_R\lambda({\bm x},t)F({\bm x},t)\,d^3x\,dt\right] \eq6
$$
is now the {\it partition functional}. The Lagrange multiplier function is identified as the solution
of the functional differential equation
$$
\langle F({\bm x},t)\rangle\equiv{\rm Tr}[\rho F({\bm x},t)]=-\frac{\delta}{\delta\lambda({\bm x},t)}\ln Z\,,
\qquad ({\bm x},t)\in R\,, \eq7
$$
and is defined only in the region $R$.
Note carefully that the data set denoted by $\langle F({\bm x},t)\rangle$
is a numerical quantity that has been equated to an expectation value to incorporate it into a
density matrix. Any other operator $J({\bm x},t)$, including $J=F$, is determined at any other space-time
point $({\bm x},t)$ as usual by
$$
\langle J({\bm x},t)\rangle={\rm Tr}\bigl[\rho J({\bm x},t)\bigr]={\rm Tr}\bigl[\rho(t)J({\bm x})\bigr]\,.
\eq8
$$
That is, the system with fixed $H$ still evolves unitarily from the initial nonequilibrium state (5); although
$\rho$ surely will no longer commute with $H$, its eigenvalues nevertheless remain unchanged.

Inclusion of a number of operators $F_k$, each with its own information-gathering region $R_k$ and
its own Lagrange multiplier function $\lambda_k$, is straightforward.
If $F$ is actually time independent an equilibrium distribution of the form (3) results, and a further
removal of spatial dependence brings us back to the canonical distribution of the original PME. But the full
form (5) illustrates how $\rho$ naturally incorporates memory effects while placing no restrictions on
spatial or temporal scales. Nor are there any issues of retardation, for example, since the procedure is a
matter of inference, not dynamics (at this point).

Some further discussion is required here. The density matrix $\rho$ in (5) is {\it not} a function of
space and time; it merely provides an initial nonequilibrium distribution corresponding to data $\langle F(
{\bm x},t)\rangle\in R$. Lack of any other information outside $R$ --- in the
future, say --- may tend to render $\rho$ less and less reliable, and the quality of predictions may
deteriorate.
The maximum entropy itself is a functional of the initial values $\langle F_k({\bm x},t)\rangle\in R_k$ and follows
from substitution of (5) into the information entropy:
$$
S_{\rm noneq}[\{\langle F_k\rangle\}]\equiv k\ln Z[\{\lambda_k\}]+k\sum_k\int_{R_k}\lambda_k({\bm x},t)
\langle F_k({\bm x},t)\rangle\,
d^3x\,dt\,.\eq9
$$
Although there is no obvious connection of $S_{\rm noneq}$ with the thermodynamic entropy, it does provide a
measure of the number of microscopic states consistent with the history of a system over the $R_k({\bm x},t)$;
it might thus be interpreted as the physical entropy of the initial nonequilibrium state (5).
If we visualize the evolution of a microstate as a path in `phase space-time', then $S_{\rm noneq}$ is the
cross section of a tube formed by all paths by which the given history could have been realized, a natural
extension of Boltzmann's $S_{\rm B}=k\ln W$, where $W$ is a measure of the set of microscopic states compatible with the
macroscopic constraints on the system. In this sense $S_{\rm noneq}$  governs the theory of irreversible
processes in much the same way as the Lagrangian governs mechanical processes. The role of entropy is thus greatly expanded
to describe not only the present nonequilibrium state, but also the recent thermokinetic history leading
to that state. We begin to see that here, unlike the equilibrium situation, entropy is intimately related to
processes.

If the information-gathering region $R$ is simply a time interval we arrive at the initial state $\rho(t_0)$
considered in the previous section. Restriction of $R$ to only a spatial region leads to a description of inhomogeneous
systems. For example, specifying the particle number density $\langle n({\bm r})\rangle$
throughout the system, in addition to $H$, constitutes a separate piece of data at each point in the volume,
and hence requires a corresponding Lagrange multiplier at each point. The distribution (3) is then replaced by
$$
\rho=\frac1{Z}\exp\left[-\beta H+ \int\lambda({\bm r}^\prime)n({\bm r}^\prime)\,d^3r^\prime\right]\,,
\eq{10}
$$
which reduces to the grand canonical distribution if $n({\bm r})$ is in fact spatially constant throughout $V$, or 
if only the volume integral of $n(\bm r)$ is specified. A similar expression is obtained if,
rather than specifying or measuring $\langle n({\bm r})\rangle$, the inhomogeneity is introduced by means of
an external field coupled to $n({\bm r})$. In that case $\lambda({\bm r})$ is given as a field strength and
$\langle n({\bm r})\rangle$ is to be determined; that is, $\lambda$ is taken
as an independent variable. Extensive application of (10) to inhomogeneous systems is given in the review by
Evans (1979).

To this point there has been no mention of dynamic time evolution; we have only described how to construct a 
single, though arbitrary, nonequilibrium macrostate based on data ranging over a space-time region. A first step 
away from this restriction is to consider steady-state systems, in which there may be currents, but all variables
are time independent. The main dynamical features of the equilibrium state are that it deals only with constants of 
the motion, among which is the density matrix itself: $[H,\rho]=0$. These constraints characterize the time-invariant
state of a closed system,
because the vanishing of the commutator implies that $\rho$ commutes with the time-evolution operator, 
so that all expectation values are constant
in time. The time-invariant state of an open system is also stationary, but $H$ almost certainly will not
commute with the operators $\{F_k\}$ defining that state, and hence not with $\rho$. Nevertheless, we can add
that constraint explicitly as the {\it definition} of a steady-state probability distribution, and
the requirement that $[\rho,H]=0$ leads to the result that
only that part of $F_k$ that is diagonal in the energy representation is included in $\rho$.\fnd{This
prescription for stationarity was advocated earlier by Fano (1957),
and has also been employed by Nakajima (1958) and by Kubo, {\it et al} (1985).}
It is reasonably straightforward to show ({\it e.g.}, Grandy, 1988) that a
representation for the diagonal part of an operator is given by
$$ \eqalign{F^d
&=F-\lim_{\epsilon\to 0^+}\int_{-\infty}^0 e^{\epsilon t}\,\partial_t F({\bm x},t)\,dt\cr
&=\lim_{\epsilon\to 0^+} \epsilon\int_{-\infty}^0 e^{\epsilon t}\,F({\bm
x},t)\,dt \cr
&=\lim_{\tau\to\infty}\frac1{\tau}\int^0_{-\tau} F({\bm x},t)\,dt\,,\cr}\eq{11}
$$
where the time dependence of $F$ is determined by (4), and
$\epsilon>0$. The second line follows from an integration by parts; the third
is essentially Abel's theorem and equates the diagonal part with a time average over the
past, which is what we might expect for a stationary process. That is, $F^d$ is that part of $F$
that remains constant under a unitary transformation generated by $H$.

Consider a number of operators $F_k({\bm x})$ defining a steady-state process.
Then the steady-state distribution $\rho_{ss}$ is simply a modification of that described by (5)
and (6):
$$
\rho_{ss} =\frac1{Z_{ss}}\exp\left[-\sum_k\int_{R_k}\lambda_k({\bm x})F^d_k({\bm
x})\,d^3x\right]\,, \eq{12} $$
where
$$
Z_{ss}[\lambda({\bm x})]={\rm Tr}\exp\left[-\sum_k\int_{R_k}\lambda_k({\bm x})F^d_k({\bm x})\,d^3x\right]
\eq{13}
$$
We illustrate some applications of these expressions further on, but note that in their full nonlinear form
they present formidable difficulties in calculations.

This last {\it caveat} suggests that we first examine small departures from equilibrium, much in the spirit of 
Eqs.(I--17)-(I--19). Suppose the equilibrium distribution to be based on expectation values
of two variables, $\langle f\rangle$ and $\langle g\rangle$,
with corresponding Lagrange multipliers $\lambda_f$, $\lambda_g$. We also suppose that no generalized work is
being done on the system, so that only `heat-like' sources may operate.
A small change from the equilibrium distribution can be characterized by small changes in the Lagrange
multipliers, which in turn will induce small variations in the expectation values. Thus,
$$
\eqalignno{\delta\langle f\rangle &=\frac{\partial\langle f\rangle}{\partial\lambda_f}\delta\lambda_f
+\frac{\partial\langle f\rangle}{\partial\lambda_g}\delta\lambda_g\,, &\hbox{(14{\rm a})} \cr
\delta\langle g\rangle &=\frac{\partial\langle g\rangle}{\partial\lambda_f}\delta\lambda_f
+\frac{\partial\langle g\rangle}{\partial\lambda_g}\delta\lambda_g\,. &\hbox{(14{\rm b})} \cr}
$$
But from (I--13b) and (I--14)) the negatives of these derivatives are just the covariances of $f$ and $g$,
$$
\eqalign{K_{fg}=K_{gf} &\equiv -\frac{\partial\langle f\rangle}
{\partial\lambda g}=-\frac{\partial\langle g\rangle}{\partial\lambda_f} \cr
&=\langle fg\rangle -\langle f\rangle\langle g\rangle\,, \cr}
\eq{15}
$$
so (14) reduce to the matrix equation
$$
\pmatrix{\delta\langle f\rangle\cr \delta\langle g\rangle\cr}=-\pmatrix{K_{ff}&K_{fg}\cr K_{gf}&K_{gg}\cr}
\pmatrix{\delta\lambda_f\cr \delta\lambda_g\cr}\,. \eq{16}
$$

In references on irreversible thermodynamics ({\it e.g.}, de Groot and Mazur, 1962)
the quantities on the left-hand side of (16)
are called fluxes, and the $(-\delta\lambda)$s are thought of as the forces that drive the system back to equilibrium.
We can thus think of the $\lambda$s as potentials that produce such forces. Linear homogeneous relations such as
(16) were presumed by Onsager (1931), but here they arise quite naturally, and in (15) we observe the celebrated Onsager
reciprocity relations.

Suppose now that we add another constraint to the maximum-entropy construction by letting $f$ be coupled to
a weak thermal source. In addition, we shall specify that $g$ is explicitly {\it not} driven, so that any
internal changes in it can only be inferred from the changes in $f$. We thus set $\delta\lambda_g=0$ in
(16) and those equations reduce to
$$
\eqalign{\delta\langle f\rangle &=-K_{ff}\delta\lambda_f=\delta Q_f\,, \cr
\delta\langle g\rangle &=-K_{gf}\delta\lambda_f\,. \cr} \eq{17}
$$
So for small variations the change in the coupled variable is essentially the source strength itself; the internal
change in $g$ is also proportional to that source strength, but modulated by the extent to which $g$ and $f$ are
correlated:
$$
\delta\langle g\rangle=\frac{K_{gf}}{K_{ff}}\delta Q_f\,, \eq{18}
$$
exhibiting what is sometimes referred to as {\it mode-mode coupling}. These expressions are precisely what one
expects from a re-maximization of the entropy subject to a small change $\delta\langle f\rangle$. For example,
if $\delta Q_f>0$ and $f$ and $g$ are positively correlated, $K_{gf}>0$, then we expect increases in the
expectation values of both quantities, as well as a corresponding increase in the maximum entropy.

Although this discussion of small departures from equilibrium is only a first step, it reinforces, and serves
as a guide to, the important role of sources in any deeper theory. It also exhibits the structure of the first
approximation, or linearization of such a theory, which is often a necessary consideration. We return
to the essential aspects of that approximation a bit later. \bigskip

\noindent {\bf 3. Sources and Thermal Driving}\medskip

We seek a description of macroscopic
nonequilibrium behavior that is generated by an arbitrary source whose precise details may be unknown.
One should be able to infer the presence of such a source from the data, and both the strength and rate of
driving of that source should be all that are required for predicting reproducible effects. Given data
 --- expectation values, say --- that vary continuously in time, we infer a source at work and
expect $\rho$ to be a definite function of time, possibly evolving principally by external means. 
In I we argued that, because all probabilities
are conditional on some kind of given information or hypothesis, $P(A_i|I)$
can change in time only if the information $I$ is changing in time, while the propositions $\{A_i\}$ are
taken as fixed. This then served as the basis for an abstract model of time-dependent probabilities.
With this insight we can see how the Gibbs algorithm might be extended to time-varying macroscopic 
systems in a straightforward manner.

As in I, information gathered in one time interval
can certainly be followed by collection in another a short time later, and can continue to be collected in
a series of such intervals, the entropy being re-maximized subject to all previous data after each interval.
Now let those intervals
become shorter and the intervals between them closer together, so that by an obvious limiting procedure
they all blend into one continuous interval whose upper endpoint is always the current moment. Thus, there
is nothing to prevent us from imagining a situation in which our information or data are continually changing
in time. A rational\'e for envisioning re-maximization to occur at every moment, rather than all at once, can
be found by again appealing to Boltzmann's expression for the entropy: $S_{\rm B}=\ln W$.
At any moment $W$ is a measure of the phase volume of all
those microstates compatible with the macroscopic constraints --- and $\ln W$ is the maximum of the information entropy at that
instant. As Boltzmann realized, this is a valid representation of  the maximized entropy
even for a nonstationary state. It is essential to understand that $W$ is a {\it number} representing the 
multiplicity of a macrostate that changes only as a result of changing external constraints. It is not
a descriptor of which microscopic arrangements are being realized by the system at the moment --- there is
no way we can ascertain that --- but only a measure of how many such states may be compatible with the
macrostate defined by those constraints. In principle we could always compute a $W$ for a set of values
of the macroscopic constraints without ever carrying out an experiment. Thus, we begin to see how an
{\it evolving} entropy can possibly be related to the time-dependent process.

There may seem to be a problem here for someone who thinks of probabilities as real physical entities, since it
might be argued that the system cannot possibly respond fast enough for $W$ to readjust its content instantaneously.
But it is not the response of the system that is at issue here; only the set of {\it possible} microstates
compatible with the present macroscopic constraints readjusts. Those potentialities always exist and need no
physical signal to be realized. A retardation problem might exist if we were trying to follow the system's changing
occupation of microstates, but we are not, because we cannot. The multiplicity $W$ does {\it not} change just 
because the microstate occupied by the system changes; in equilibrium those changes go on continuously, but $W$ 
remains essentially constant. Only variations in the macroscopic constraints can change $W$, and those are 
instantaneous and lead to immediate change in the maximum information entropy $S_{\rm B}$.

To introduce the notion of a general source let us consider a generic system described by a density matrix
$$
\rho=\frac1{Z}e^{\alpha A+\beta B+\biggamma C}\,, \eq{19}
$$
and a process that drives the variable $B$ such that an amount $\Delta B$ is transferred into the system.
That is, $B$ is driven by some means other than dynamically, with no obvious effective Hamiltonian. In addition,
the variable $A$ is explicitly {\it not} driven, but can change only as a result of changes in $B$
if $A$ and $B$ are correlated. Since
there is no new information regarding $A$, even though it is free to readjust when $B$ is changed, the
Lagrange multiplier $\alpha$ must remain unchanged. We also add the further constraint on the process that
$C$ is to remain unchanged under transfer of $\Delta B$. This is a generalization of the scenario described by
(16), and can be summarized as follows:
$$
\eqalign{\delta\alpha &=0\,, \qquad \langle A\rangle\to \langle A\rangle^\prime\,, \cr
\delta\beta\ &\not= 0\,, \qquad \langle B\rangle\to \langle B\rangle +\Delta B\,, \cr
\delta\biggamma &=-\frac{K_{CB}}{K_{CC}}\delta\beta\,, \qquad \langle C\rangle\to \langle C\rangle\,.\cr} \eq{20}
$$
This is the most general form of a constrained driven process, except for inclusion of a number of
variables of each kind. Any such driving not tied to a specific dynamic term in a Hamiltonian will be referred
to as thermal driving.
A variable, and therefore the system itself, is said to be {\it thermally driven} if no new variables other than
those constrained experimentally are needed to characterize the resulting state, and if the Lagrange
multipliers corresponding to variables other than those specified remain constant. As discussed in I, a major
difference with purely dynamic driving is that the thermally-driven density matrix is not constrained to evolve
by unitary transformation alone.

Let us suppose that the system is in thermal equilibrium with time-independent
Hamiltonian in the past, and then at $t=0$ a
source is turned on smoothly and specified to run continuously, as described by its effect on the expectation value
$\langle F(t)\rangle$. That is, $F(t)$ is given throughout the changing interval $[0,t]$ and is
specified to continue to change in a known way until further notice.\fnd{The lower limit of the driving interval
is chosen as 0 only for convenience.} Although any complete theory of nonequilibrium must be a continuum field
theory,
we shall omit spatial dependence explicitly here in the interest of clarity and return to address that point
later. For convenience we consider only a single driven operator; multiple operators, both driven and constrained,
are readily included. Based on the probability model of I, the PME then provides the density matrix for thermal
driving:
$$
\eqalign{\rho_t &=\frac1{Z_t}\exp\left[-\beta H-\int_0^t\lambda(t^\prime)
F(t^\prime)\,dt^\prime\right]\,, \cr
Z_t[\beta,\lambda(t)] &=\hbox{\rm Tr}\,\exp\left[-\beta H-\int_0^t
\lambda(t^\prime)F(t^\prime)\,dt^\prime\right]\,, \cr} \eq{21}
$$
and the Lagrange-multiplier function is formally obtained from
$$
\langle F(t)\rangle_t=-\frac{\delta}{\delta\lambda(t)}\ln Z_t\,, \eq{22}
$$
for $t$ in the driving interval.
Reference to the equilibrium state is made explicit not only because it provides
a measure of how far the system is removed from equilibrium, but also because it removes all
uncertainty as to the previous history of the system prior to introduction of the external source; 
clearly, these are not essential features of the construction.

Since $\rho_t$ can now be considered an explicit function of $t$, we can employ the operator identity $\partial_x
e^{A(x)}=e^{A(x)}{\overline{\partial_xA}}$ to compute the time derivative:
$$
\partial_t\rho_t=\rho_t\lambda(t)
\bigl[\langle F(t)\rangle_t
-{\overline{F(t)}}\bigr]\,,
\eq{23}
$$
where the overline denotes a generalized {\it Kubo transform} with respect to the operator $\ln\rho_t$:
$$
{\overline {F(t)}}\equiv \int_0^1 e^{-u\ln\rho_t}\,F(t)e^{u\ln\rho_t}\,du\,, \eq{24}
$$
which arises here from the possible noncommutativity of $F(t)$ with itself at different times.

The expression (23) has the {\it form}
of what is often called a `master equation', but it has an entirely different origin and is exact; it is, in fact, 
the $\partial_t\rho$ term in the equation of motion (I--40). Because $\lambda(t)$ is defined
only on the information-gathering interval $[0,t]$, Eq.(23) just specifies the rate at which $\rho_t$ is
changing in that interval. Although $\rho_t$ does not evolve under time-independent $H$ in the Heisenberg
picture, in this case it does evolve explicitly, and in the Schr\"odinger picture this time variation will be in
addition to the canonical time evolution. In turn, an unambiguous time dependence for the
entropy is implied, as follows.

The theoretical maximum entropy $S_t=-k\hbox{\rm Tr}[\rho_t\ln\rho_t]$ is obtained explicitly by
substitution from (21),
$$
\frac1{k}S_t=\ln Z_t+\beta\langle H\,\rangle_t +\int_0^t \lambda(t^\prime)
\langle F(t^\prime)\rangle_t\, dt^\prime\,;
\eq{25}
$$
it is the continuously re-maximized information entropy.
Equation (25) indicates explicitly that $\langle H\,\rangle_t$ changes only as a result of changes in,
and correlation with $F$. The constraint that $H$ is explicitly not driven implies that $\langle H\,\rangle_t$
and $\langle F(t^\prime)\rangle_t$ are no longer independent, and that means that $\lambda(t)$ cannot
be determined directly from $S_t$ by functional differentiation in (25); this has important consequences.

The expectation value of another operator at time $t$ is $\langle C\,\rangle_t=\hbox{\rm Tr}[\rho_t C]$, and
direct differentiation yields
$$
\eqalign{\frac{d}{dt}\langle C(t)\rangle_t &
=\hbox{\rm Tr}\bigl[C(t)\partial_t\rho_t +\rho_t{\dot{C}}(t)\bigr] \cr
&=\langle{\dot{C}}(t)\rangle_t -\lambda(t)
K_{CF}^t(t,t)\,,\cr} \eq{26}
$$
where the superposed dot denotes a total time derivative.
We have here introduced the {\it covariance function}
$$
K_{CF}^t(t^\prime,t)\equiv \langle{\overline{F(t^\prime)}}
C(t)\rangle_t-\langle F(t^\prime)\rangle_t\langle C(t)\rangle_t=
-\frac{\delta\langle C(t)\rangle_t}{\delta\lambda(t)}\,,  \eq{27}
$$
which is a quantum mechanical generalization of the covariance (15).
Note that all of the preceding entities are completely nonlinear, in that expectation values, Kubo transforms,
and covariance functions are all written in terms of the density matrix $\rho_t$, which is the meaning of the
superscript $t$ on $K_{CF}^t$.  Although time-translation invariance is
not a property of the general nonequilibrium system, it is not difficult to show that the reciprocity relation
$K^t_{CF}(t^\prime,t)=K^t_{FC}(t,t^\prime)$ is valid.

Let us introduce a new notation into (26), which at first appears to be only a convenience:
$$
{\bigsigma}_C(t)\equiv \frac{d}{dt}\langle C(t)\rangle_t-\langle {\dot C}(t)\rangle_t
=-\lambda(t)K_{CF}^t(t,t)\,.
\eq{28}
$$
For a number of choices of $C$ and $F$ the equal-time covariance function vanishes,
but if $C=F$ an illuminating interpretation first noticed by Mitchell (1967) emerges:
$$
\eqalign{{\bigsigma}_F(t) &\equiv \frac{d}{dt}\langle F(t)\rangle_t
-\langle {\dot F}(t)\rangle_t  \cr
&=-\lambda(t)K_{FF}^t(t,t)\,.\cr}
\eq{29}
$$
Owing to the specification of thermal driving, $d\langle F(t)\rangle_t/dt$ is the total time rate-of-change
of $\langle F(t)\rangle_t$ in the system at time $t$, whereas $\langle {\dot F}(t)\rangle_t$
is the rate
of change produced by internal relaxation. Hence, ${\bigsigma}_F(t)$ must be the rate at which $F$ is
driven or transferred
by the external source, and is often what is measured or controlled experimentally. One need know nothing else
about the details of the source, because its total effect on the system is expressed by the second equality
in (29). If the source strength is given, then (29) is a nonlinear transcendental equation determining the Lagrange
multiplier function $\lambda(t)$.

An important reason for eventually including spatial dependence is that we can now {\it derive} the
macroscopic equations of motion. For example, if $F(t)$ is
one of the conserved densities $e({\bm x},t)$ in a simple fluid and ${\bm J}({\bm x},t)$ the corresponding
current density, then the local microscopic continuity equation
$$
{\dot e}({\bm x},t)+\boldnabla\cdot{\bm J}({\bm x},t)=0 \eq{30}
$$
is satisfied irrespective of the the state of the system. When this is substituted into (29) we obtain
the macroscopic conservation law
$$
\frac{d}{dt}\langle e({\bm x},t)\rangle_t +\boldnabla\cdot\langle{\bm J}({\bm x},t)\rangle_t=
{\bigsigma}_e({\bm x},t)\,, \eq{31}
$$
which is completely nonlinear. Specification of sources therefore provides automatically the thermokinetic
equations of motion; for example, if $e$ is the momentum density $m{\bm j}({\bm x},t)$, so that $\bm J$ is
the stress tensor $T_{ik}$, then a series of transformations turns (31) into
the Navier-Stokes equations of fluid dynamics.

\heading
Nonequilibrium Thermodynamics
\endheading

The notion of thermal driving provides a basis for nonequilibrium thermodynamics, which can be
developed in much the same way as is done for the equilibrium theory ({\it e.g.}, Grandy, 1987).
As with that case, the operator $F$ can also depend on an external variable $\alpha$, so that at time
$t$ the entropy is $S_t=S_t[\langle H\rangle_t, \langle F(t)\rangle_t; \alpha]$; of course, we could also
include a number of other measured variables $\{F_i\}$, though only $H$ and $F$ will be employed here.
But now $S_t$ is also a function of time and, from (25), its total time derivative is
$$
\frac1{k}\frac{dS_t}{dt}=\left(\frac{\partial\ln Z_t}{\partial\alpha}\right){\dot\alpha}
+\beta\frac{d\langle H\rangle_t}{dt}
-\lambda(t)\int_0^t\lambda(t^\prime)K^t_{FF}(t,t^\prime)\,dt^\prime\,. \eq{32}
$$
Although $\partial_t Z_t$ contributes to ${\dot S}_t$, its contribution is cancelled because
$$
\partial_t\ln Z_t =-\lambda(t)\langle F(t)\rangle_t\,, \eq{33}
$$
which also provides a novel representation for $Z_t$ upon integration.
In principle, then, one can follow
the increase (or decrease) of entropy in the presence of external sources (or sinks).

The most common type of external variable $\alpha$ is the system volume $V$, so that in the 
equilibrium theory $(\partial\langle H\rangle/\partial V)dV=-P\,dV$ is an element of work. This 
suggests a general interpretation of the first term on the right-hand side of (32). As an example,
in the present scenario consider the simple process of an adiabatic free expansion of
a gas, wherein only the work term is involved in (32). We can now model this by specifying a form
for $\alpha=V$; for example, $V(t)=V_0\bigl(2-e^{-bt}\bigr)$ would, for $b$ very large, rapidly
inflate the volume
to double its size over an interval from $t=0$ to some later time $\tau$. One also needs
an equation of state for the gas, but usually the pressure is proportional to $V^{-1}$ and therefore
decreases exponentially as well. In the case of an ideal gas, integration of this form for ${\dot S}_t$
over ($0,\tau$) yields the expected change $S_{\tau}-S_0=kN\ln 2$. This result is almost independent of
the model as long as $V(\tau)\simeq 2V_0$.

Ordinarily $\dot\alpha=0$. In this case we can also explicitly evaluate the term containing the Hamiltonian
and rewrite (31) as
$$\eqalign{
\frac1{k}\frac{dS_t}{dt} &=-\beta\lambda(t)K^t_{HF}(t,0)
-\lambda(t)\int_0^t
\lambda(t^\prime)K^t_{FF}(t,t^\prime)\,dt^\prime\cr
&=\biggamma_F(t){\bigsigma}_F(t)\,, \cr} \eq{34}
$$
where we have employed (29) and defined a new parameter
$$
\biggamma_F(t)\equiv\beta\frac{K^t_{HF}(t,0)}{K^t_{FF}(t,t)}
+\int_0^t\lambda(t^\prime)\frac{K^t_{FF}(t,t^\prime)}{K^t_{FF}(t,t)}\,dt^\prime\,. \eq{35}
$$

Although this expression for $\biggamma$ at first glance seems only a bookkeeping convenience, it is
actually of some physical significance, as suggested by (20). As noted above, the
thermal driving constraint on $H$
prevents  $\langle H\rangle_t$  and $\langle F(t)\rangle_t$ from being completely independent; indeed,
neither of them is independent of $\langle F(t^\prime)\rangle_{t}$.
In turn, and unlike the equilibrium case,
$\partial\langle f_m\rangle/\partial\lambda_n$ and $\partial\lambda_n/ \partial\langle f_m\rangle$ are
no longer the respective elements of a pair of mutually inverse matrices. Thus, $\delta S_t/\delta
\langle F(t)\rangle_t$ does not determine $\lambda(t)$; rather, from (25),
$$
\frac{\delta S_t}{\delta\langle F(t)\rangle_t}=\frac{\delta\langle H\rangle_t}{\delta\lambda(t)}
\frac{\delta\lambda(t)}{\delta\langle F(t)\rangle_t} +\int_0^t\lambda(t^\prime)\frac{\delta\langle F(t^\prime)
\rangle_{t}}{\delta\lambda(t)}\frac{\delta\lambda(t)}{\delta\langle F(t)\rangle_t}\,dt^\prime\,.\eq{36}
$$
Owing to interdependencies we can now write $\delta\lambda(t)/\delta\langle F(t)\rangle_t=1/K^t_{FF}(t,t)$,
and hence the right-hand side of (36) is just $\biggamma_F(t)$, which now has the general definition
$$
\biggamma_F(t)\equiv \left(\frac{\delta S_t}{\delta\langle
F(t)\rangle_t}\right)_{{\rm thermal}\atop {\rm driving}}\,. \eq{37}
$$
The subscript ``thermal driving" reminds us that this derivative is somewhat different than its usual form
in equilibrium.
When the source strength $\bigsigma_F(t)$ is specified the Lagrange multiplier itself is determined from (29).

Physically, $\biggamma_F$ is a {\it transfer potential} in the same sense that the $\lambda$s in Eq.(16)
are thought of as potentials. Just as products of potentials and expectation values appear in the structure
of the equilibrium entropy, in thermal driving the entropy
production (34) is always a sum of products of transfer potentials and source terms measuring the
rate of transfer.
So, the entropy production is not in general given by products of `fluxes' and `forces', and $S_t$ and $\dot S_t$
are not simple generalizations of equilibrium quantities.
But the ordinary potentials also play another role in equilibrium: if two systems in contact can exchange
energy and particles, then they are in equilibrium if the temperatures and chemical potentials of the two
are equal. Similarly, if two systems can exchange  quantities $F_i$ under thermal driving, then the conditions for
{\it migrational equilibrium} at time $t$ are
$$
\biggamma_{F_i}(t)_1=\biggamma_{F_i}(t)_2\,. \eq{38}
$$
Migrational equilibrium in stationary processes is discussed, for example, by Tykodi (1967).

What is the physical interpretation to be given to $S_t$? Clearly it refers only to the information
encoded in the distribution of (21) and cannot refer to the internal entropy of the system. In equilibrium
the maximum of this information entropy is the same as the experimental entropy, but that is not necessarily 
the case here.
For example, if the driving is removed at time $t=t_1$, then $S_{t_1}$ in (25) can only provide the entropy
of that nonequilibrium state; its value will remain the same during subsequent relaxation, owing to unitary
time evolution. Although the maximum information (or theoretical) entropy provides a complete description of the system
based on all known physical constraints on that system, it cannot describe the ensuing relaxation, for it contains
no new information about that process. Nevertheless, $S_t$  does have a
definite physical interpretation.

The form of $\bigsigma_F$ in (29) suggests a natural separation of the entropy if that expression is
substituted into the second line of (34):
$$
{\dot S}_t=\biggamma_F(t)\left(\frac{d}{dt}\langle F(t)\rangle_t -\langle{\dot F}(t)\rangle_t\right)\,. \eq{39}
$$
Thus, ${\dot S}_t$ has the qualitative form ${\dot Q}/T$, as intuition might have suggested.
The first term on the right-hand side must represent the total time rate-of-change of entropy  ${\dot S}_{tot}$
arising from
the thermal driving of $F(t)$, whereas the second term is the rate-of-change of internal entropy ${\dot S}_{int}$
owing to relaxation. Thus, the total entropy production can be written
$$
{\dot S}_{tot}(t)={\dot S}_t +{\dot S}_{int}(t)\,, \eq{40}
$$
where the {\it entropy production of transfer} owing to the external source, ${\dot S}_t$, is given by (34).
This latter
quantity is a function only of the driven variable $F(t)$, whereas the internal entropy depends on all variables, driven or not,
necessary to describe the nonequilibrium state and is determined by the various relaxation
processes taking place in the system. Calculation of ${\dot S}_{int}$, of course, depends on a rather detailed model
of the system; we'll
have more to say on this below.\fnd{Equation (40) is reminiscent of similar expressions for entropy changes, 
such as $dS = dS_{ext} + dS_{int}$,
that can be found in works on phenomenological nonequilibrium thermodynamics ({\it e.g.}, de Groot and Mazur, 1962).}

In an equilibrium system the major role of $S$ is associated with the Second Law, and
this law in its traditional form has little to say about nonequilibrium processes. In these
latter processes, however, it is ${\dot S_t}$, rather than $S_t$ itself that plays the major role, as is seen
in (34)-(37). That is, ${\dot S_t}$ governs the transfer process in terms of the rate of driving and
the transfer potential, in much the same way that $S$ governs the direction of changes between equilibrium
states through $dQ/T$. In nonequilibrium processes ${\dot S_t}$ also governs the rate; this is true even in the
steady state when one takes into account sources and sinks.

The distinction between theoretical entropy in equilibrium scenarios and in nonequilibrium processes cannot be
emphasized enough. If external forces are removed, it is a mathematical theorem that
neither $\rho_t$ nor $S_t$ can evolve into their equilibrium counterparts. This is a singular limit, 
as discussed earlier, and unless these distinctions are  clearly recognized
few real advances can be made in nonequilibrium statistical mechanics.\bigskip

\heading
Constant Driving Rate and Spatial Variation
\endheading

To complete the general development, logical consistency requires an examination of thermal driving at a constant rate.
For this purpose it will first be useful to record the generalizations of the primary equations
of thermal driving to include spatial coordinates:
$$
\partial_t\rho_t = \rho_t\int\lambda({\bm x}^\prime,t)\bigl[\langle F({\bm x}^\prime,t)\rangle_t
-\overline{F({\bm x}^\prime,t)}\bigr]\,d^3x^\prime\,, \eq{41}
$$
$$
\frac1{k}S_t =\ln Z_t +\beta\langle H\rangle_t +\int d^3x^\prime\int^t_0dt^\prime\, \lambda({\bm x}^\prime,
t^\prime)\langle F({\bm x}^\prime,t^\prime)\rangle_t\,, \eq{42}
$$
$$
\eqalign{\frac1{k}{\dot S}_t &=-\beta\int \lambda({\bm x}^\prime,t)K_{HF}^t({\bm x}^\prime,t)\,d^3x^\prime \cr
&\qquad-\int d^3x^{\prime\prime}\lambda({\bm x}^{\prime\prime},t)\int d^3x^\prime\int_0^t dt^\prime
\lambda({\bm x}^{\prime\prime},t^\prime)K_{FF}^t({\bm x}^{\prime\prime},t;{\bm x}^\prime,t^\prime)\,,\cr}
\eq{43}
$$
$$
\sigma_F({\bm x},t) = -\int \lambda({\bm x}^\prime,t)K_{FF}^t({\bm x}^\prime,t;{\bm x},t)\,. \eq{44}
$$

This last expression can be inverted by introducing an inverse integral operator:
$$
\lambda({\bm x},t) =-\int\bigl[K^t_{FF}({\bm x}^\prime,t;{\bm x},t)\bigr]^{-1}\sigma_F({\bm x}^\prime,t)\,
d^3x^\prime\,, \eq{45}
$$
which is a nonlinear integral equation for $\lambda({\bm x},t)$. Thus, the right-hand side of (45) is
really only a shorthand notation for the iterated solution.  Upon
substitution of (45) into (43) we find that
$$
\frac1{k}{\dot S}_t =\int\biggamma_F({\bm x},t)\sigma_F({\bm x},t)\,d^3x\,, \eq{46}
$$
where
$$
\eqalign{\biggamma_F({\bm x},t) &\equiv \beta\int d^3x^\prime\bigl[K^t_{FF}({\bm x},t;{\bm x}^\prime,t)\bigr]^{-1}
K^t_{HF}({\bm x}^\prime,t)\cr
&\qquad\int d^3x^\prime\int d^3x^{\prime\prime}\int^t_0dt^\prime\lambda({\bm x}^{\prime\prime},t)
\bigl[K^t_{FF}({\bm x},t;{\bm x}^\prime,t)\bigr]^{-1}K_{FF}^t({\bm x}^\prime,t;{\bm x}^{\prime\prime},t^\prime)\,.
\cr} \eq{47}
$$

We can verify this expression for $\biggamma_F$ from the more general definition
$$
\biggamma_F({\bm x},t)\equiv \left(\frac{\delta S_t}{\delta\langle F({\bm x},t)\rangle_t}
\right)_{{\rm thermal}\atop {\rm driving}}\,, \eq{48}
$$
if we note two properties of functional differentiation. First, the ordinary chain rule for partial
differentiation of $F[x(s),y(s)]$ with respect to s,
$$
\frac{\partial F}{\partial s}=\frac{\partial F}{\partial x}\frac{\partial x}{\partial s}+
\frac{\partial F}{\partial y}\frac{\partial y}{\partial s}\,,
$$
generalizes to
$$
\eqalign{\frac{\delta\langle G({\bm x},t)\rangle}{\delta\langle F({\bm x},t)\rangle}
&=\int\frac{\delta\langle G({\bm x},t)\rangle}{\delta\lambda_F({\bm x}^\prime,t)}
\frac{\delta\langle\lambda_F({\bm x}^\prime,t)}{\delta\langle F({\bm x},t)\rangle}\,d^3x^\prime \cr
&=\int K^t_{GF}({\bm x}^\prime,t;{\bm x},t)\bigl[K^t_{FF}({\bm x},t;{\bm x}^\prime,t)\bigr]^{-1}
\,d^3x^\prime\,,\cr}\eq{49}
$$
for example. Second, in Eq.(42) for $S_t$ the upper limit $t$ on the time integral, and the subscript
on $\langle H\rangle_t$, prevent the functional derivative from yielding merely $\lambda({\bm x},t)$, which
is determined by $\sigma_F({\bm x},t)$ at any rate. Rather, we obtain (47) for $\biggamma_F({\bm x},t)$.

Specification of constant driving means that $\sigma_F$ is constant in time, and from (29) or (44) this in
turn implies that $\lambda({\bm x},t)$ must actually be independent of time. This last assertion follows
because the covariance function in these equations is time independent, owing to the re-emergence of unitary
time evolution in the absence of internal time variation. That is, the integrals in (21), generalized
to include spatial variables, can now be rewritten in the form
$$
\int d^3x^\prime \lambda({\bm x}^\prime)\int_0^t F({\bm x}^\prime,t^\prime)\,dt^\prime\,. \eq{50}
$$
But now the form of the time integral no longer makes sense in the context of time-independent driving.

If a constant rate of driving is specified as a constraint on the initial probability distribution we
take this to mean that the initial data were constant in the distant past, and at least up to the time
of observation. In requiring this one faces the possibility of a divergent integral, so that it is
necessary to {\it regularize} the integral, along the lines of methods often employed in quantum field
theory. In the present case we rewrite the time integral in (50) as a time average over the past:
$$
\lim_{\tau\to\infty}\frac1{\tau}\int_{-\tau}^0 F({\bm x}^\prime,t^\prime)\,dt^\prime\,. \eq{51}
$$
This, however, is just the diagonal part of the operator $F({\bm x}^\prime)$ as given by Eq.(11),
and hence constant driving corresponds with our definition of the steady state. In this scenario we can then
replace all the time integrations over operators by the diagonal parts of those operators and omit all
time dependence. We see that, in the sense of this procedure, the steady state is also a singular limit
of the general nonequilibrium state, in that the latter does not reduce in a completely straightforward
mathematical way to the former.

In the steady state we expect time derivatives of all expectation values to vanish; hence from (29) we
have the further implication that
the constant rate of driving is exactly balanced by the rate of internal relaxation. This is how the system
responds to steady currents.

Although there exist stationary currents within the system, the steady driving takes place in the terminal
parts, or boundaries of the system, and such currents imply irreversible dissipation. There must then be
an overall rate of dissipation or entropy production generated by the external sources. This rate is
provided by Eq.(46), now rewritten in the form
$$
\frac1{k}{\dot S}_t =\int\biggamma_F({\bm x})\sigma_F({\bm x})\,d^3x\,. \eq{52}
$$
The general definition of $\biggamma_F({\bm x})$ still applies, but the explicit form is now
$$
\eqalign{\biggamma_F({\bm x}) &\equiv \beta\int d^3x^\prime\bigl[K^{ss}_{F^dF^d}({\bm x};{\bm x}^\prime)\bigr]^{-1}
K^{ss}_{HF^d}({\bm x}^\prime)\cr
&\qquad\int d^3x^\prime\int d^3x^{\prime\prime}\lambda({\bm x}^{\prime\prime})
\bigl[K^{ss}_{F^dF^d}({\bm x};{\bm x}^\prime)\bigr]^{-1}K_{F^dF^d}^{ss}({\bm x}^\prime,;{\bm x}^{\prime\prime})\,.
\cr} \eq{53}
$$ \bigskip

\noindent {\bf 4. The Linear Approximation}\medskip

Much, though not all, of the work on macroscopic nonequilibrium phenomena has of necessity centered on
small departures from equilibrium, or the linear approximation, so that it is of some value to outline
that reduction of the present theory and discuss briefly some applications.
We envision situations in which the system has been in thermal equilibrium in the remote past and later
found to produce data of the form considered above. By considering both pieces of data we obtain a measure
of the departure from equilibrium. In describing the general method of linearization
the character of the perturbing term and the scenario under consideration are immaterial; hence, we 
can take the distribution (21) with integration limits replaced by the space-time region $R$ as our
generic model and, for brevity, temporarily omit space dependences.\fnd{The equilibrium distribution is taken as canonical
only for convenience; for example, one could just as well use the grand canonical form, as well as include
different types of particle.} Thus, we consider the model
$$
\eqalignno{\rho &=\frac1{Z}\exp\biggl\{-\beta H-\int_R \lambda(t)F(t)\,dt\biggr\}\,, &\hbox{(54)} \cr
\noalign{\vskip 3pt}
Z[\beta,\lambda(t)] &={\rm Tr}\exp\biggl\{-\beta H-\int_R \lambda(t)F(t)\,dt\biggr\}\,, &\hbox{(55)} \cr}
$$
where $\beta$ refers to the temperature of the previous equilibrium state --- no other value of $\beta$ makes sense
until the system returns to equilibrium.

By linear approximation we mean ``linear in the departure from equilibrium." In the present case
that means that the entire integral in (54) and (55) is in some sense small. An expansion of the
exponential operator follows from repeated application of the identity
$$
e^{A+B}=e^A\biggl[1+\int_0^1 e^{-xA}\,Be^{x(A+B)}\,dx\biggr]\,, \eq{56}
$$
where $B$ is the small perturbation.
The first-order, or linear approximation to the expectation value of another operator $C$ is (Heims and
Jaynes, 1962; Jaynes, 1979; Grandy, 1988)
$$
\langle C\rangle\simeq\langle C\rangle_0 -\int_0^1\big\langle
e^{-xA}Be^{xA}\,C\bigr\rangle_0\,dx+ \langle B\rangle_0\langle C\rangle_0\,,
\eq{57} 
$$
where $\langle B\rangle_0={\rm Tr}\bigl(e^A\,B\bigr)$.
In (57) we again encounter the Kubo transform of the operator $B$ with respect to $A$, the nonlinear form of
which was introduced in (24).

Application of this approximation scheme to (54) and (55) reveals that the leading-order departure
of the expectation value of $C$ at time $t$ from its equilibrium value is
$$
\langle C(t)\rangle-\langle C\rangle_0=-\int_R K_{CF}(t,t^\prime)\lambda(t^\prime)\, dt^\prime\,, \eq{58}
$$
where $K_{CF}\equiv K^0_{CF}$ is the linearized version of the covariance function defined in (27):
$$
\eqalign{K_{CF}(t,t^\prime) &\equiv\langle{\overline{F(t^\prime)}} C(t)\rangle_0-\langle
F\rangle_0\langle C\rangle_0 \cr
&=-\frac{\delta\langle C(t)\rangle}{\delta\lambda(t)}\,, \cr} \eq{59}
$$
and $\langle\cdots\rangle_0$ is an expectation value in terms of the equilibrium distribution $\rho_0$.
Time independence of the Hamiltonian confers the same property upon the single-operator expectations, and also
guarantees time-translation invariance: $K_{CF}(t,t^\prime)=K_{CF}(t-t^\prime)$. One verifies the reciprocity relation
$$
K_{CF}(t-t^\prime)=K_{FC}(t^\prime-t)\, \eq{60}
$$
from a change of variables and cyclic invariance of the trace. Note that it is always the second variable that carries 
the Kubo transform. If $C$ and $F$ are Hermitian, $K_{CF}$ is real and $K_{FF}\ge 0$. In this case
$K_{CF}$ has all the properties of a scalar product on a linear vector space, and thus satisfies
the Schwarz inequality:
$K_{CC}K_{FF}-K_{CF}^2\ge 0$, with equality if and only if $C=cF$, with $c$ a real constant.

The covariance function (59) clearly depends only on equilibrium properties of the system. Quite
generally, then, small departures from equilibrium caused by {\it anything} are described principally by
equilibrium fluctuations. While this provides some useful physical insight, the
other side of the coin is that covariance functions are exceedingly difficult
to calculate for interacting particles, other than in some kind of
perturbation theory. The linear approximation represents considerable
progress, but formidable mathematical barriers remain.  In practice, however, it is usually the relations among these and other
quantities that interest us; after all, we seldom evaluate from first principles the derivatives
in the Maxwell relations, yet they provide us with important insights. Linear hydrodynamics
provides one area in which various approximation schemes for correlation functions have proved
fruitful.

While the Lagrange multiplier function $\lambda(t)$ is determined formally by (7),
one suspects that if we set $C=F$ and restrict $t$ to the region $R$, then (58) becomes a Fredholm
integral equation determining $\lambda(t)$ in the only interval in which it is defined. This indeed turns out to
be the case, though the demonstration that the two procedures are equivalent requires a little effort
(Grandy, 1988). This is, in fact, a very rich result, and to discuss it in slightly more detail it will be convenient to
specify $R$ more definitely, as $[-\tau,0]$, say. Thus, the  expression
$$
\langle F(t)\rangle-\langle F\rangle_0=-\int_{-\tau}^0
K_{FF}(t-t^\prime)\lambda(t^\prime)\,dt^\prime \eq{61}
$$
is now seen to have several interpretations as $t$ ranges over $(-\infty,\infty)$. When $t>0$ it gives the
predicted future of $F(t)$; with $-\tau\le t\le 0$ it provides a linear integral equation determining $\lambda(t)$;
and when $t<-\tau$ it yields the retrodicted past of $F(t)$. This last observation underscores the facts
that $K_{FF}(t)$ is not necessarily a causal function unless required to be so, and that these expressions are
based on probable inference; in physical applications
the dynamics enters into computation of the covariance function, but does not dictate
its interpretation in various time domains. Although physical influences must propagate
forward in time, logical inferences about the present can affect our knowledge of the past as well as the future.
Retrodiction, of course, is at the heart of fields such as aarcheology cosmology, geology, and paleontology.

When the perturbed system is
spatially nonuniform we find that (58) and (59) are replaced by
$$\eqalignno{\langle C({\bf x},t)\rangle-\langle C({\bf x})
\rangle_0 &=-\int_R K_{CF}({\bf x},t;{\bf x}^{\prime},t^{\prime})
\lambda({\bf x}^{\prime},t^{\prime})\,
d^3x^{\prime}\,dt^{\prime}\,, &\hbox{(62)}\cr
K_{CF}({\bf x},t;{\bf x}^{\prime},t^{\prime}) &=\langle\overline{
F({\bf x}^{\prime},t^{\prime})}C({\bf x},t)\rangle_0 -
\langle F({\bf x}^{\prime})\rangle_0\langle C({\bf x})
\rangle_0\,, &\hbox{(63)}\cr}
$$
so that in its causal domain $K_{CF}({\bf x},t;{\bf x}^{\prime},t^{\prime})$ takes the
form of a Green function.
Note that the single-operator expectation values are also independent
of ${\bf x}$ in an initially homogeneous system, and that the generalization to include
a number of operators $F_k({\bf x},t)$ is straightforward.

If the equilibrium system is also space-translation invariant it is useful to
employ the notation ${\bm r}\equiv {\bm x}-{\bm x}^\prime$.
Generally, the operators encountered in covariance functions possess definite transformation
properties under space inversion (parity) and time reversal. Under the former $A({\bm r},\tau)$
becomes $P_AA(-{\bm r},\tau),\,P_A=\pm 1$, and under the latter $T_AA({\bm r},-\tau),\,T_A=\pm 1$.
For operators describing a simple fluid, say , $PT=+1$ and one verifies that the full reciprocity relation holds:
$$
K_{CF}({\bm r},\tau)=K_{FC}({\bm r},\tau)\,. \eq{64}
$$
The efficacy of these equations of the linear approximation will become apparent as we present some sample
applications.

\heading
Linear Transport Processes
\endheading

The generic model for a macroscopic fluid is most readily described as a continuum in terms of various densities,
and representations in terms of quantum-mechanical operators are defined in terms of field operators in a
Fock representation ({\it e.g.},
Fetter and Walecka, 1971). The three basic density operators in the fluid are the number density $n$, momentum density
$m{\bm j}$, and energy density $h$, where $\bm j$ is the particle current-density operator.
Unless so specified, these generally have no explicit
time dependence, so that their equations of motion in the Heisenberg picture are
$$
\eqalignno{{\dot n}({\bm x},t) &=\frac{i}{\hbar}[H,n({\bm x},t)]\,, &\hbox{(65{\rm a})}\cr
m{\dot{\bm j}}({\bm x},t) &=\frac{i}{\hbar}[H,m{\bm j}({\bm x},t)]\,, &\hbox{(65{\rm b})}\cr
{\dot h}({\bm x},t) &=\frac{i}{\hbar}[H,h({\bm x},t)]\,. &\hbox{(65{\rm c})}\cr}
$$
But the left-hand sides of these equations are also involved in statements of the local microscopic conservation
laws in the continuum, which usually relate time derivatives of densities to divergences of the corresponding currents.
The differential conservation laws are thus obtained by evaluating the commutators on the right-hand sides in the forms
$$
\eqalignno{{\dot n}({\bm x},t) &=-\boldnabla\cdot{\bm j}({\bm x},t)\,, &\hbox{(66{\rm a})}\cr
m{\dot {\bm j}}({\bm x},t) &=-\boldnabla\cdot{\bm T}({\bm x},t)\,, &\hbox{(66{\rm b})}\cr
{\dot h}({\bm x},t) &=-\boldnabla\cdot{\bm q}({\bm x},t)\,. &\hbox{(66{\rm c})}\cr}
$$
The superposed dot in these equations indicates a total time derivative. In the absence of external
forces and sources (65) are equivalent to unitary transformations, and the Hamiltonian
and total-number operator, respectively, are given by
$$
H=\int h({\bm x},t)\,d^3x\,, \qquad N=\int n({\bm x},t)\,d^3x\,, \eq{67}
$$
both independent of time.

The current density $\bm j$ is just the usual quantum-mechanical probability current density, so that (66a) is
easily verified. Identification of the energy current density $\bm q$ and stress tensor $\bm T$, however, is far from
straightforward; in fact, they may not be uniquely defined for arbitrary particle-particle interactions. But if the
Hamiltonian is rotationally invariant we can restrict the discussion to
spherically-symmetric two-body potentials. Two further symmetry properties
arise from time independence and spatial uniformity in the equilibrium system:
time-translation and space-translation invariance, respectively. These latter
two invariances are expressed in terms of volume-integrated, or total energy,
number, and momentum operators, so that the commutators $[H,{\bm P}]$,
$[H,N]$, $[{\bm P},N]$ all vanish. Specification of these symmetry properties
defines a {\it simple fluid}, and the operators $\bm q$ and $\bm T$ can be
identified uniquely by evaluation of the commutators in Eqs.(65b,c). 
The algebra is tedious and the results are given, for example, by
Puff and Gillis (1968), and Grandy (1988). Thus, the five local microscopic conservation
laws (66) completely characterize the simple fluid and lead to five
long-lived hydrodynamic modes. Local disturbances of these quantities cannot
be dissipated locally, but must spread out over the entire system.

As a first application of the linear theory we return to the steady-state scenario of Eqs.(12) and (13)
and also incorporate a term $-\beta H$ in the exponentials to characterize an earlier equilibrium
reference state.
Denoting the deviation from equilibrium as $\Delta F({\bm x})=
F({\bm x})-\langle F({\bm x})\rangle_0$, we find that in linear approximation another operator $C$ will have
expectation value
$$
\eqalign{\langle\Delta C({\bm x})\rangle_{ss} &=-\int_R\lambda({\bm x}^\prime)K_{CF}({\bm x}-{\bm x}^\prime)\,
d^3x^\prime \cr
&\qquad +\lim_{\epsilon\to 0^+}\int_R d^3x^\prime\,\int_{-\infty}^0 e^{\epsilon t}\,\lambda({\bm x}^\prime)
K_{C\dot F}({\bm x}-{\bm x}^\prime,t)\,dt\,, \cr} \eq{68}
$$
where we have employed the expression (11) for the diagonal part of an operator, and subscripts $ss$ refer 
to the steady-state distribution. Specify
$F({\bm x})$ to be one of the fluid densities $d({\bm x})$, so that the continuity equations (66) lead to
the identity
$$
\frac{d}{dt}K_{dB}({\bm x},t)=-\boldnabla\cdot K_{{\bm j}B}({\bm x},t)\,, \eq{69}
$$
and thus $K_{C\dot d}$ in (68) can be replaced by $-\boldnabla^\prime\cdot
K_{C{\bm J}}$. Let $R({\bm x})$ be the system volume $V$, and presume
$K_{C{\bm J}}$ to vanish at large distances. An integration by parts then reduces (68) to
$$
\eqalign{\langle\Delta C({\bm x})\rangle_{ss} &=-\int_V\lambda({\bm x}^\prime)K_{Cd}({\bm x}-{\bm x}^\prime)\,
d^3x^\prime \cr
&\qquad +\lim_{\epsilon\to 0^+}\int_V d^3x^\prime\,\int_{-\infty}^0 e^{\epsilon t}\,
\nabla^\prime\lambda({\bm x}^\prime)\cdot
K_{C{\bm J}}({\bm x}-{\bm x}^\prime,t)\,dt\,, \cr} \eq{70}
$$
in which we have dropped the surface term.

Classical hydrodynamics corresponds to a long-wavelength approximation by presuming that
$\nabla^\prime\lambda$ varies so slowly that it is effectively constant over the range for
which $K_{C{\bm J}}$ is appreciable.\fnd{We presume that the fluctuations are not correlated over the
entire volume.} With this in mind we can extract the gradient from the
integral and write 
$$
\eqalign{\langle\Delta C({\bm x})\rangle &\simeq -\int_V\lambda({\bm x}^\prime)K_{Cd}({\bm x}-{\bm x}^\prime)\,
d^3x^\prime \cr
&\qquad +\nabla\lambda\cdot\lim_{\epsilon\to 0^+}\int_v d^3x^\prime\,\int_{-\infty}^0
e^{\epsilon t}\,K_{C{\bm J}}({\bm x}-{\bm x}^\prime,t)\,dt\,, \cr} \eq{71}
$$
which is the fundamental equation describing linear transport processes in the steady state.
The integration region $v$ is the correlation volume, ooutsideof which the correlations vanish; it is 
introduced here simply as a reminder that the spatial correlations are presumed to be of short range.

As an example, let $d$ be the number density $n$  with gradient characterized by the deviation $\Delta n({\bm x})=n({\bm x})-
\langle n\rangle_0$. The specified density gradient and the predicted current density, respectively, are then
$$
\eqalignno{\langle\Delta n({\bm x})\rangle &=-\int_V\lambda({\bm x}^\prime)K_{nn}({\bm x}-{\bm x}^\prime)\,
d^3x^\prime \cr
&\qquad +\nabla\lambda\cdot\int_v d^3x^\prime\,\int_{-\infty}^0
e^{\epsilon t}\,K_{n{\bm j}}({\bm x}-{\bm x}^\prime,t)\,dt\,  \cr
&=-\int_V\lambda({\bm x}^\prime)K_{nn}({\bm x}-{\bm x}^\prime)\,
d^3x^\prime\,,  &\hbox{(72)}\cr}    % \noalign{\vskip 3pt}
$$
$$
\eqalignno{\langle{\bm j}({\bm x})\rangle &=-\int_V\lambda({\bm x}^\prime)K_{{\bm j}n}({\bm x}-{\bm x}^\prime)\,
d^3x^\prime \cr
&\qquad +\nabla\lambda\cdot\int_v d^3x^\prime\,\int_{-\infty}^0
e^{\epsilon t}\,K_{{\bm j}{\bm j}}({\bm x}-{\bm x}^\prime,t)\,dt\,  \cr
&= \nabla\lambda\cdot\int_v d^3x^\prime\,\int_{-\infty}^0
e^{\epsilon t}\,K_{{\bm j}{\bm j}}({\bm x}-{\bm x}^\prime,t)\,dt\,, &\hbox{(73)}\cr }
$$
where the limit $\epsilon\to 0^+$ is understood.
We have noted that the second term of the first line in (72)
and the first term of the first line in (73) vanish by symmetry.

Now take the gradient in (72), make the long-wavelength approximation, and eliminate $\nabla\lambda$ 
between this result and (73), which leads to the relation
$$
\eqalign{\langle {\bm j}({\bm x})\rangle &=-\frac{\int_0^\infty e^{-\epsilon t}\,dt\,\int_v K_{{\bm j}{\bm j}}
({\bm x}-{\bm x}^\prime,t)\,d^3x^\prime}{\int_v K_{nn}({\bm x}-{\bm x}^\prime)\,d^3x^\prime}\cdot\nabla\langle
n({\bm x})\rangle \cr
&\equiv -{\bm D}({\bm x})\cdot\nabla\langle n({\bm x})\rangle\,, \cr} \eq{74}
$$
with the proviso that $\epsilon\to 0^+$.
This is {\it Fick's law of diffusion}, in which we have identified the
diffusion tensor $\bm D$ that can now be calculated in principle from
microscopic dynamics; owing to spatial uniformity in the equilibrium system $D({\bm x})$
is actually independent of $\bm x$. For more general nonequilibrium states the same type of
calculation produces a quantity ${\bm D}({\bm x},t)$ having the same form as that in (74), and
the long-wavelength approximation also involves one of short memory. (By `short memory' we mean that
recent information is the most relevant, not that the system somehow forgets.)

It is remarkable that linear constitutive equations such as Fick's law arise from almost nothing more
than having some kind of data available over a space-time region.
These relations have long been
characterized as phenomenological, since they
are not derived from dynamical laws. We now see why this is so, for the derivation here shows that they
are actually {\it laws of of inference}. Indeed, what we usually mean by `phenomenological' is `inferred from experience',
a notion here put on a sound footing through probability theory. When they are coupled with the
corresponding conservation laws, however, one does obtain macroscopic dynamical laws, such as the
diffusion equation.

Because it involves a slightly different procedure, and will provide a further example below, let us
consider thermal conductivity (which need not be restricted to fluids). A steady gradient in energy density
is specified in the form of a deviation $\Delta h({\bm x})=h({\bm x})-\langle h\rangle_0$. By a calculation
similar to the above we find for the expected steady-state heat current
$$
\langle {\bm q}({\bm x})\rangle=\int_v d^3x^\prime\int_0^\infty e^{-\epsilon t}\,\nabla\lambda({\bm x}^\prime)
\cdot K_{\bm qq}({\bm x}-{\bm x}^\prime, t)\,dt\,, \eq{75}
$$
where the limit $\epsilon\to 0^+$ is understood, and we have not yet invoked the long-wavelength limit.
In this case we do not eliminate $\nabla\lambda$, for it contains the gradient of interest. Both
dimensionally, and as dictated by the physical scenario, $\lambda$ must be $\beta({\bm x})=[kT({\bm x})]^{-1}$,
a space-dependent temperature function. Although such a quantity may be difficult to measure in general,
it is well-defined in the steady state. With this substitution the long-wavelength approximation of
constant temperature gradient in (75) yields
$$
\eqalign{\langle {\bm q}({\bm x})\rangle &\simeq -\nabla T\cdot\int_v d^3x^\prime\int_0^\infty e^{-\epsilon t}\,
\frac{K_{\bm qq}({\bm x}-{\bm x}^\prime, t)}{kT^2({\bm x}^\prime)}\,dt  \cr
&\equiv -{\bm\kappa}\cdot\nabla T({\bm x})\,, \cr} \eq{76}
$$
in which we identify the thermal conductivity tensor $\bm\kappa$, which again is independent of $\bm x$.
This is
{\it Fourier's law of thermal conductivity}; it applies to solids as well as fluids, but calculation
of the covariance function remains a challenge. It is left to the reader to verify that $\bm\kappa$, as well
as $\bm D$ in (74), are positive.

A common model employing (76) is that of a uniform conducting rod of length $L$
and thermal conductivity $\bm \kappa$. We can calculate the constant rate of transfer of entropy from the
source to the sink by means of (52), in which the transfer potential $\biggamma({\bm x})$
is simply the spatial temperature
distribtution $\beta({\bm x})$, and $\sigma({\bm x})$ is the (constant) rate of driving on the end boundaries
of the rod. In this case the driving rate is given by the heat current $\langle {\bm q}\rangle$ itself,
inserting thermal energy at one end and taking it out at the other. Hence,
$$
\eqalign{\frac1{k}{\dot S}_t &=\int_0^L\frac1{kT(x)}(-{\bm\kappa}\nabla T)\bigl[\delta_{x,0}-\delta_{x,L}\bigr]\,dx\cr
&=\frac{{\bm\kappa}}{L}\frac{\bigl(T_H-T_C\bigr)^2}{T_H T_C}\,,\cr} \eq{77}
$$
which is identical to the more intuitively obtained result ({\it e.g.}, Palffy-Muhoray, 2001).
Although (52) itself is completely
nonlinear, one notes that we have employed the linear form of Fourier's law (76) for the current.
This calculation illustrates the
importance of boundary conditions in describing stationary processes; Tykodi (1967) has also emphasized the
role of {\it terminal parts} in describing the steady state.

\heading
Linear Response Theory
\endheading

An important feature of the thermal driving mechanism is that the actual details of the thermal driving source
are irrelevant, and only the {\it rates and strengths} at which system variables are driven enter the equations.
It should make no difference in many situations whether the driving is thermal or mechanical; we
examine the latter context here.

The theory of dynamical response was described very briefly in Eqs.(I--5)-(I--8), and the linear version follows as
described there. The underlying scenario is that a well-defined external field is imposed on a system that has
been in thermal equilibrium in the remote past, as described by the Hamiltonian $H_0$.
It is then
presumed that the response to this disturbance can be derived by adding a time-dependent term to the
Hamiltonian, so that effectively $H=H_0-Fv(t)$, $t>0$, where $v(t)$ describes the external field and $F$ is a system
operator to which it couples. Some of the difficulties with this approach were sketched in I, including the
observation that $\rho(t)$ can only evolve unitarily. We now see that these problems can be resolved by noting
that dynamical response is just a special case of thermal driving. 

For eventual comparison with the results of linear response theory we shall need an identity for the time derivative
of the covariance function. Direct calculation in the definition (59) yields
$$
\eqalign{\frac{d}{d\tau}K_{CF}(\tau) &= \frac{i}{\beta\hbar}\bigl\langle[C, F(\tau)]\bigr\rangle_0 \cr
&=-\beta^{-1}\phi_{CF}(\tau)\,, \cr} \eq{78}
$$
where $\phi_{CF}$ is the linear response function. Clearly, the covariance function contains a good deal more
information than does the dynamic response function.

The derivation of the generic maximum-entropy distribution in (I--14) disguises a subtle point regarding that
procedure. We note from (I--16) that the Lagrange multiplier $\lambda$ can also be determined from the maximum 
entropy: 
$$
\lambda=\frac1{k}\frac{\partial S}{\partial\langle f\rangle}\,. \eq{79}
$$
Together with (I--15) this reveals a reciprocity implying that the probability distribution can be obtained
by specifying {\it either} $\langle f\rangle$ {\it or} $\lambda$.
An example of this choice is illustrated in the canonical distribution
(I--10), which could be obtained by specifying either the energy or the temperature; this option
was also exercised in the model of spatial inhomogeneity of Eq.(10). Thus, we return to Eqs.(21),
replacing $H$ with $H_0$, and let $\lambda(t^\prime)$ be the independent variable. In linear approximation (29)
expresses $\lambda(t)$ directly in terms of the source strength, or driving rate, and dimensional considerations 
suggest that we write this variable in the form 
$$
\eqalign{\lambda(t^\prime) &=\beta\frac{d}{dt^\prime}\bigl[\theta(t-t^\prime)v(t^\prime)\bigr] \cr
&=\beta\left[-\delta(t-t^\prime)v(t^\prime)+\theta(t-t^\prime)\frac{d}{dt^\prime}v(t^\prime)\right]\,, \cr} \eq{80}
$$
with the condition that $v(0)=0$. The step-function $\theta(t-t^\prime)$ is included in (80) because $\lambda$
is defined {\it only} on the interval [$0,t$].

Substitution of (80) into (21) yields the distribution relevant to a well-defined external field,
$$
\eqalign{\rho_t &=\frac1{Z_t}\exp\left[-\beta H_0+\beta\int_0^t\Bigl[\delta(t-t^\prime)-\theta(t-t^\prime)\frac{d}{dt^\prime}
\Bigr]v(t^\prime) F(t^\prime)\,dt^\prime\right] \cr
&=\frac1{Z_t}\exp\left[-\beta H_0 +\beta\int_0^tv(t^\prime){\dot F}(t^\prime)\,dt^\prime\right]\,, \cr} \eq{81}
$$
and $Z_t$, as usual, is the trace of the numerator.
Although the exponential contains what appears to be an effective Hamiltonian, we do {\it not} assert that
$\int_0^t v(t^\prime){\dot F}(t^\prime)\,dt^\prime$ is an addition to the equilibrium Hamiltonian $H_0$;
there is no rational\'e of any kind for such an assertion. The
Lagrange multiplier function $\lambda(t)$ is a macroscopic quantity, as is its expression as an independent variable
in (80). The linear approximation (58), along with the identity ((78), yields the departure from equilibrium
of the expected value of another operator $C$ at any future time $t$ under driving by the external field:
$$
\eqalign{\langle C(t)\rangle-\langle C\rangle_0 &=\beta\int_0^t v(t^\prime)K_{C{\dot F}}(t-t^\prime)\,dt^\prime \cr
&= \beta\int_0^t v(t^\prime)\frac{d}{dt^\prime}K_{CF}(t-t^\prime)\,dt^\prime \cr
&=\int_0^t v(t^\prime)\phi_{CF}(t-t^\prime)\,dt^\prime\,, \cr} \eq{82}
$$
which is precisely the result obtained in linear response theory. But now we also have the time-evolved probability
distribution (81) from which we can develop the associated thermodynamics.
Equation (82) confirms that, at least linearly, both $\rho_t$ and a unitarily evolved $\rho(t)$ will predict the 
same expectation values. But, as suggested following (78), $\rho(t)$ contains no more macroscopic information 
than it had to begin with.

As an example of an
external source producing a time-varying field, suppose a component of electric polarization
$M_i(t)$ is specified, leading to the  density matrix
$$
\rho_t=\frac1{Z_t}\exp\left[-\beta H_0 +\int_0^t\lambda_i(t^\prime)M_i(t^\prime\,dt^\prime\right]\,.
\eq{83}
$$
We presume no spontaneous polarization, so that in linear approximation the expectation of another
component at time $t$ is
$$
\langle M_j(t)\rangle=\int_0^t\lambda(t^\prime)\langle{\overline{M_i(t^\prime)}}M_j(t)\rangle_0\,dt^\prime\,.
\eq{84}
$$

Now, with the additional knowledge that (84) is the result of turning on an external field
one might be led to think that the
Lagrange multiplier is simply a field component, say $E_i(t)$. But (80) shows that, even when the
{\it effect} is to add a time-dependent term to the Hamiltonian, the actual source term is somewhat more
complicated; only the $\delta$-function term in (80) corresponds to that possibility, and the actual
source term also describes the rate of change of the field. This again illustrates the earlier observation
that the covariance function contains much more information than the dynamic response function.

With (80) we can rewrite (84) explicitly as
$$
\eqalign{\langle M_j(t)\rangle =\beta\langle{\overline{M_i(t)}} &M_j(t)\rangle_0 E_i(t) \cr
&-\beta\int_0^t \langle{\overline{M_i(t^\prime)}}M_j(t)\rangle_0\frac{dE_i(t^\prime)}{dt^\prime}\,
dt^\prime\,, \cr} \eq{85}
$$
which is just the result obtained from the theory of dynamic response. But we've
uncovered much more, because now one can do thermodynamics. In the present scenario we have
specified thermal driving of the polarization and incorporated that
into a density matrix; additionally, the Lagrange multiplier has been chosen to be the independent variable
corresponding to an external field, which allows us to identify the source strength.
Thus, we have a definite expression for the time-dependent entropy of the ensuing
nonequilibrium state:
$$
\eqalign{\frac1{k} S_t &=\ln Z_t +\beta\langle
H\rangle_t-\int_0^t\lambda(t^\prime)\langle M(t^\prime)\rangle_t \,dt^\prime\,, \cr
&\simeq\frac1{k} S_0 +\beta\int_0^t\lambda(t^\prime)K_{H_0M}(t^\prime)\,dt^\prime +{\rm O}(\lambda)\,, \cr} \eq{86}
$$
where the second line is the linear approximation and we have identified the entropy of the equilibrium system as 
$S_0=k\ln Z_0+k\beta\langle H_0\rangle_0$. In the case of dynamic response, if one makes the linear approximation
to $\rho(t)$ in (I--6) and computes the entropy similarly, it is found that $S(t)-S_0$ vanishes identically, as
expected. With (86), however, the entropy difference can also be written in terms of the linear response function:
$$
\frac1{k}\bigl(S_t-S_0\bigr)\simeq \beta\int_0^t v(t^\prime)\phi_{H_0M}(t^\prime)\,dt^\prime\,. \eq{87}
$$
These remarks strongly suggest that the proper theory of response to a dynamical perturbation is to be
found as a special case of thermal driving.

\heading
Relaxation
\endheading

When external sources are removed we expect the system to relax to a (possibly new) state of thermal equilibrium.
If the driving ceases at time $t=t_1$, say, then from that point on the system is
described by (21) with the replacement $t\to t_1$ everywhere, barring any further external influence. These equations define
the nonequilibrium state at $t=t_1$, from which the subsequent behavior can be predicted.

As discussed earlier,  $S_{t_1}$ as given by (25) cannot evolve to the entropy of some equilibrium state,
for the same reason that $\rho_{t_1}$
cannot evolve to a canonical equilibrium distribution; both evolve from $t=t_1$ under unitary transformation. It
should be sufficient, however, to show that the macrovariables describing the thermodynamic system, such as
$\langle F(t)\rangle_{t_1}$, may relax to a set of equilibrium values. Then, with those predicted values, we can
construct a new canonical density matrix via entropy maximization that will describe the new equilibrium state. 
The value $S_{t_1}$ remains the entropy of the nonequilibrium state at the time the driving was removed.

Calculation of the exact expectation values is essentially intractable, of course, so we again employ the linear
approximation. For example, at time $t\ge t_1$ the expectation of $F(t)$ itself is
$$
\eqalign{\Delta F(t) \equiv\langle F(t)\rangle_{t_1} - \langle F\rangle_0 &\simeq\int_0^{t_1}\lambda(t^\prime)K_{FF}(t-t^\prime)\,
dt^\prime \cr
&= \int_0^{t_1}\sigma_F(t^\prime)\frac{K_{FF}(t-t^\prime)}{K_{FF}(0)}\,dt^\prime\,, \cr} \eq{88}
$$
where we've utilized (29). Although everything on the right-hand side of (88) is presumably known, we see
that the actual details of the relaxation process depend crucially on the behavior of $K_{FF}(t-t^\prime)$ for
$t>t_1$.

In the discussion following Eq.(60) we noted that the covariance functions satisfy the Schwarz inequality. From this we
can see that the ratio $r(t-t^\prime)=|K_{FF}(t-t^\prime)/K_{FF}(0)|$ in (88), and therefore the integrand, reach their maxima at
$t^\prime=t$ where $r(0)=1$. Further, $r(t-t^\prime)$ is less than unity for $t^\prime < t$, and again for all $t>t_1$; 
the exact magnitude of $r$ depends on the decay properties of $K_{FF}(t-t^\prime)$. In any event, the major contribution
to the integral arises from the region around the cutoff $t=t_1$. The {\it relaxation time} $\tau$ can be estimated 
by studying the asymptotic properties of the time-derivative of $\Delta F(t)$, which in turn requires an examination of the
time-derivative of $K_{FF}(t-t^\prime)$. From (78), we are thus seeking a time $t_2$ for which
$$
\left|\langle[F(t^\prime),F(t)]\rangle_0\right| \ll 1\,, \eq{89}
$$
by some criterion. Then, $\tau\simeq t_2-t_1$. 

For many covariance functions the ratio $r(t-t^\prime)$ in (88) will tend to some constant value as $t/t_1$
becomes large, while others may tend to zero. For example, if we turn off the burner under a pot of water at $t=t_1$, the
total energy of the equilibrium system will be $\langle E(t_1)\rangle_{t_1}$, so that $K_{EE}$ would be expected to reach its
nonzero asymptotic form very quickly. But in the polarization example of (84) we expect the correlations to decay to zero
as the system relaxes back to the unpolarized state; this may, or may not, be rapid. One can only uncover the particular
behavior from a detailed study of the covariance functions as determined by the relaxation mechanisms specific
to the system, which are generally governed by particle interactions.

A final point we can make here concerns the rate-of-change of internal entropy, ${\dot S}_{int}$. When the source is
removed  the entropy production of transfer immediately vanishes for $t>t_1$. Equation (40) then implies that the
total rate of entropy production is entirely that of relaxation, and the compelling conclusion is that ${\dot S}_{int}$
is actually the {\it relaxation rate} and may be observable. This interpretation should also be valid when ${\dot S}_t\not= 0$,
which is strongly indicated in the steady state scenario.\bigskip

\noindent {\bf 5. Summary}\medskip

The aim of this discussion has been to expand the concept of theoretical entropy in equilibrium thermodynamics to
encompass macroscopic systems evolving in time. In doing so we find that the maximum information entropy $S_t$, while 
providing a complete description of the nonequilibrium state at any instant, does not assume the dominant role it does in
an equilibrium context. Rather, the rates and directions of processes are the most important features of 
nonequilibrium systems, and the rate of entropy production ${\dot S}_t$ takes the form of a transfer potential times
a rate of transfer, or a generalized intuitive form of ${\dot Q}/T$. This suggests that in nonequilibrium thermodynamics 
it is ${\dot S}_t$ that governs the ongoing macroscopic processes and can be expressed as a measurable quantity
via Eq.(34). In the absence of external sources (or sinks) the rate of entropy production simply describes the 
relaxation rate; the theoretical maximum entropy itself characterizes only the nonequilibrium state from which the 
system is relaxing to the singular equilibrium limit. Further thought leads us to conclude that these interpretations can
also be applied to ongoing processes.

The formalism presented here applies to macroscopic systems arbitrarily far from equilibrium, although the nonlinear 
equations provide formidable mathematical barriers to any detailed calculations. While the linear approximation is the 
most fruitful approach, even here the covariance functions remain somewhat complicated and resistant to exact 
computation; various attacks have produced some progress, however, in the context of linear hydrodynamics. At 
present it is the formal relations containing covariance functions that can prove most useful, in which 
carefully chosen models of these nonequilibrium correlations can play a role similar to that of potential models in
equilibrium statistical mechanics. Although the present results may lay some groundwork for a complete theory of 
nonequilibrium thermodynamics, there is a great deal of room for expansion and further development. 

\bigskip\medskip
\centerline{\bf REFERENCES}
 \bigskip

\nh Berry, M. (2002), ``Singular Limits," Physics Today {\bf 55} (5), 10.

\nh Clausius, R. (1865), ``\"Uber verschiedene f\"ur die Anwendung bequeme Formen der Hauptgleichungen
der mechanische W\"armetheorie," Ann. d. Phys.[2] {\bf 125}, 390.

\nh de Groot, S.R. and P. Mazur (1962), {\it Nonequilibrium Thermodynamics}, North-Holland, Amsterdam.

\nh Evans, R. (1979), ``The nature of the liquid-vapour interface and other
topics in the statistical mechanics of non-uniform, classical fluids," Adv.
Phys. {\bf 28}, 143.

\nh Fano, U. (1957), ``Description of States in Quantum Mechanics by Density
Matrix and Operator Techniques," Rev. Mod. Phys. {\bf 29}, 74.

\nh Fetter, A.L. and J.D. Walecka (1971), {\it Quantum Theory of Many-Particle
Systems}, McGraw-Hill, New York.

\nh Gibbs, J.W. (1902), {\it Elementary Principles in Statistical
Mechanics}, Yale University Press, New Haven, Conn.

\nh Grandy, W.T., Jr.(1987), {\it Foundations of Statistical Mechanics, Vol.I: Equilibrium Theory},
Reidel, Dordrecht.

\nh\r (1988), {\it Foundations of Statistical Mechanics, Vol.II:
Nonequilibrium Phenomena}, Reidel, Dordrecht.

\nh\r (2003), ``Time Evolution in Macroscopic Systems. I: Equations of Motion," preceding paper.

\nh Jaynes, E.T. (1963), ``Information Theory and Statistical Mechanics," in K.W. Ford (ed.), {\it Statistical 
Physics}, Benjamin, New York.

\nh\r (1967), ``Foundations of Probability Theory and Statistical Mechanics,"
in M. Bunge (ed.), {\it Delaware Seminar in the Foundations of Physics},
Springer-Verlag, New York.

\nh\r (1979), ``Where Do We Stand On Maximum Entropy?," in R.D.Levine and M.
Tribus (eds.), {\it The Maximum Entropy Formalism}, M.I.T. Press, Cambridge,
MA.

\nh Kubo, R., M. Toda, and N. Hashitsume (1985), {\it Statistical Physics II}, Springer-Verlag, Berlin.

\nh Mitchell, W.C. (1967), ``Statistical Mechanics of Thermally Driven
Systems," Ph.D. thesis, Washington University, St. Louis (unpublished).

\nh Nakajima, S. (1958), ``On Quantum Theory of Transport Phenomena," Prog.
Theor. Phys. {\bf 20}, 948.

\nh Onsager, L. (1931), ``Reciprocal Relations in Irreversible Processes. I,"
Phys. Rev. {\bf 37}, 405.

\nh Puff, R.D. and N.S. Gillis (1968), ``Fluctuations and Transport Properties
of Many-Particle Systems," Ann. Phys. (N.Y.) {\bf 6}, 364.

\nh Tykodi, R.J. (1967), {\it Thermodynamics of Steady States}, Macmillan, New York.

\bye